\documentclass[conference]{IEEEconf}

\input epsf
\usepackage{graphicx}
\usepackage{multirow}
\usepackage{titlesec}
\usepackage{balance} 
\usepackage{stfloats}
\usepackage{xcolor}
\usepackage{booktabs}
\usepackage{subcaption}
\usepackage{caption}
\usepackage{url}

\renewcommand\thesection{\arabic{section}} % arabic numerals for the sections

\renewcommand\thesubsectiondis{\thesection.\arabic{subsection}}% arabic numerals for the subsections

\renewcommand\thesubsubsectiondis{\thesubsectiondis.\arabic{subsubsection}}% arabic numerals for the subsubsections

\begin{document}

\title{\textbf{\Large A comparative analysis of automated techniques for security bug report identification\\}}

\author{Muhammad Laiq\\
	\normalsize Blekinge Institute of Technology, Karlskrona, Sweden\\
	\normalsize muhammad.laiq@bth.se\\
	%\normalsize *corresponding author
}
%+++++++++++++++++++++++++++++++++++++++++++

%\specialpapernotice{(Invited Paper)}

\maketitle
\begin{abstract}
Timely identification of security-related bug reports is essential to minimize the window of vulnerabilities in software systems. Manually screening incoming bug reports to identify security-related issues is time-consuming, error-prone, and non-scalable for large-scale software systems. Thus, a variety of automatic techniques, including traditional machine learning (ML) techniques and large language models, have been proposed to facilitate this task. However, the literature remains fragmented. Most studies introduce or optimize a particular technique and evaluate it against a limited set of baselines, often under different experimental setups. As a result, it is difficult to compare their results and draw reliable conclusions about the effectiveness of existing approaches, leaving researchers and practitioners without clear guidance on which techniques are most suitable for the task.

To address this gap, we conducted a comparative analysis of several promising automated techniques to identify security-related bug reports using benchmark datasets. We evaluated Logistic Regression, Support Vector Machines, Random Forest, OpenAI's GPT-5.2, BERT-base, RoBERTa, and SetFit (a state-of-the-art few-shot learning framework).

Our results indicate that SetFit achieves the best overall performance, achieving an F1-score of 0.80 and outperforming other techniques on three of the four datasets. RoBERTa performs competitively and approaches SetFit in some projects, while traditional ML techniques, particularly Logistic Regression, remain a strong baseline in certain contexts. In contrast, GPT-5.2 performs poorly in both zero-shot and few-shot settings. In addition, cross-project experiments demonstrate that transfer learning can improve performance for projects with limited data, but may degrade results for projects with strong project-specific characteristics.

%In general, our findings highlight the challenges faced by both traditional ML techniques and LLMs under severe class imbalance and limited labeled data, while demonstrating the suitability of few-shot learning, specifically SetFit, for identifying security bug reports.
\end{abstract}
\IEEEoverridecommandlockouts
\vspace{1.5ex}

% for peerreview papers, inserts a page break and creates the second title.
% Will be ignored for other modes.
\IEEEpeerreviewmaketitle

\section{Introduction}

Software organizations rely on bug tracking systems, such as Bugzilla\footnote{\url{https://www.bugzilla.org/}} and Jira\footnote{\url{https://www.atlassian.com/software/jira}}, to manage bug reports throughout the software development lifecycle. These systems receive bug reports from various stakeholders, including developers, testers, and end-users, and typically encompass a wide range of quality concerns such as performance degradation, functional failures, stability issues, and security vulnerabilities. Among these categories, security-related bug reports are particularly critical. In contrast to non-security issues, security vulnerabilities have the potential to expose systems to unauthorized access, data breaches, or service disruption. Consequently, the timely and accurate identification of security-related bug reports is essential to prevent exploitation and mitigate risk in software ecosystems.

Manually inspecting large volumes of incoming bug reports to identify potential security issues is a laborious and error-prone process \cite{wu2021data}. Contemporary software projects, especially those that are community-driven or large-scale, may receive hundreds or even thousands of bug submissions, making the manual process increasingly infeasible. Thus, to assist practitioners in identifying security-related bug reports, several automated techniques have been proposed in the literature, for example, \cite{peters2017text,zhou2017automated,goseva2018identification,shu2019better,francca2025gpts}. Most of these approaches rely on traditional machine learning (ML) techniques, such as Logistic Regression and Random Forest, to classify bug reports into security-related and non-security-related bug reports based on textual features. More recently, researchers have begun exploring the use of more advanced large language models (LLMs), which offer enhanced capabilities for understanding contextual and semantic information in natural language \cite{yokoyama2024identifying,francca2025gpts,laiq2026few}.

Despite this growing body of work and the increasing diversity of proposed techniques, the literature remains fragmented and lacks a systematic comparison of existing automated techniques for identifying security-related bug reports. Prior studies typically focus on proposing or refining individual techniques and evaluating them against a limited set of baselines, often under differing datasets and experimental setups. This makes it difficult to compare the results of these studies and to draw reliable conclusions about the effectiveness of these techniques, leaving researchers unable to determine which methodological advances are truly effective and practitioners without clear guidance on which techniques are most suitable for adoption in practice.

To address the aforementioned gap, this study conducted a comparative analysis of several promising automated techniques to identify security-related bug reports using benchmark datasets (Ambari, Camel, Wicket, and Derby). We systematically evaluated traditional ML techniques (Logistic Regression, Support Vector Machines, and Random Forest), BERT-based LLMs (BERT-base and RoBERTa), a state-of-the-art few-shot learning framework (SetFit), and a GPT-based LLM (GPT-5.2). The contributions of this study are as follows.

\begin{itemize}
    \item We summarize existing work on automatic identification of security-related bug reports.
    \item We comprehensively evaluate several promising automatic techniques under a consistent experimental setup on benchmark datasets.
    \item We investigate transfer learning through cross-project prediction to assess when training on other projects helps or hurts performance.
    \item We investigate the impact of the commonly used data imbalance handling technique (i.e., Synthetic Minority Over-sampling Technique \cite{chawla2002smote}) on the prediction accuracy of classical ML techniques.
    \item We explore and compare prompt engineering strategies for the GPT-based model, specifically zero-shot, few-shot with two examples, and few-shot using 1\% of the training data.
    \item We evaluate RoBERTa (an additional) technique for identifying security bug reports and compare it with existing approaches.
\end{itemize}

This paper is organized as follows. Section \ref{sec:relatedwork} describes the related work on automatic identification of security-related bug reports. Section \ref{sec:Method} presents the research method of the study. Section \ref{sec:Results} presents the results of the study. Section \ref{sec:Discussion} discusses the findings of the study. Section \ref{sec:validitythreats} describes the validity threats to the study and, finally, Section \ref{sec:Conclusions} concludes the paper.

\section{Related work} \label{sec:relatedwork}
\begin{table}[!ht]
    %\centering
    \small
    \caption{An overview of the techniques, features, evaluation strategies, and evaluation metrics used in previous work}
    \label{tab:related-work-on-BRC}
    \begin{tabular}{p{4cm}p{2.8cm}r}
    \toprule

\textbf{Techniques} & \textbf{Studies} & \textbf{Count} \\ \midrule
Random Forest & \cite{laiq2026few,francca2025gpts,soltaniani2025security,cao2022sbrpbert,yokoyama2024identifying,shu2019better,peters2017text,jiang2020ltrwes,zhou2017automated} & 9 \\
Logistic Regression & \cite{laiq2026few,francca2025gpts,shu2019better,peters2017text,jiang2020ltrwes,zhou2017automated} & 6 \\
Support Vector Machines & \cite{laiq2026few,francca2025gpts,zou2018automatically,jiang2020ltrwes,zhou2017automated} & 5 \\
K-Nearest Neighbor & \cite{shu2019better,peters2017text,jiang2020ltrwes,zhou2017automated} & 4 \\
Naive Bayes & \cite{shu2019better,peters2017text,jiang2020ltrwes,zhou2017automated} & 4 \\
Multilayer Perceptron & \cite{shu2019better,peters2017text,jiang2020ltrwes} & 3 \\
BERT-base & \cite{soltaniani2025security,cao2022sbrpbert,yokoyama2024identifying} & 3 \\
Fasttext & \cite{alqahtani2024security} & 1 \\
SetFit & \cite{laiq2026few} & 1 \\ 
CNN and BiLSTM  & \cite{cao2022sbrpbert} & 1 \\
AdaBoost, Gradient boosting, and Stacking-based Ensemble & \cite{zhou2017automated} & 1 \\
ChatGPT\textbf{*} & \cite{yokoyama2024identifying} & 1 \\
GPT-based models: GPT4All-Falcon, Instruct, Open-Orca (Mistral), and Wizard & \cite{francca2025gpts} & 1 \\ \midrule

\textbf{Features} & \textbf{Studies} & \textbf{Count} \\
\midrule
Description of a bug report & \cite{laiq2026few,francca2025gpts,soltaniani2025security,cao2022sbrpbert,yokoyama2024identifying,alqahtani2024security,shu2019better,peters2017text,jiang2020ltrwes,zhou2017automated} & 10 \\
Summary  of a bug report & \cite{zou2018automatically,soltaniani2025security,alqahtani2024security,shu2019better,peters2017text,jiang2020ltrwes} & 6 \\
Title/heading of a bug report & \cite{laiq2026few,cao2022sbrpbert,zhou2017automated} & 3 \\
Meta features, such as reported time, priority, and creation time. & \cite{zou2018automatically,jiang2020ltrwes,zhou2017automated} & 3 \\ 
Commit messages and comments & \cite{zhou2017automated} & 1 \\
\midrule

\textbf{Evaluation strategies} & \textbf{Studies} & \textbf{Count}  \\ \midrule
Test on a fixed set & \cite{zou2018automatically,soltaniani2025security,cao2022sbrpbert,shu2019better,peters2017text,jiang2020ltrwes} & 6 \\
10-fold cross-validation & \cite{alqahtani2024security} & 1 \\
3-fold cross-validation & \cite{yokoyama2024identifying} & 1 \\
10 repetitions & \cite{francca2025gpts} & 1 \\
5-fold cross-validation & \cite{laiq2026few} & 1 \\
10-fold cross-validation & \cite{zhou2017automated} & 1 \\
TemplateStickiness & \cite{francca2025gpts} & 1 \\
\midrule

\textbf{Evaluation metrics} & \textbf{Studies} & \textbf{Count} \\ \midrule
Precision & \cite{laiq2026few,francca2025gpts,zou2018automatically,soltaniani2025security,cao2022sbrpbert,yokoyama2024identifying,alqahtani2024security,shu2019better,peters2017text,zhou2017automated} & 10 \\
Recall & \cite{laiq2026few,francca2025gpts,zou2018automatically,soltaniani2025security,cao2022sbrpbert,yokoyama2024identifying,alqahtani2024security,shu2019better,jiang2020ltrwes,zhou2017automated} & 10 \\
F -- Score & \cite{laiq2026few,francca2025gpts,zou2018automatically,soltaniani2025security,cao2022sbrpbert,yokoyama2024identifying,alqahtani2024security,peters2017text,jiang2020ltrwes} & 9 \\
G-measure & \cite{soltaniani2025security,alqahtani2024security,shu2019better,peters2017text,jiang2020ltrwes} & 5 \\
Accuracy & \cite{francca2025gpts,zou2018automatically,cao2022sbrpbert,yokoyama2024identifying} & 4 \\
False Positive Rate & \cite{soltaniani2025security,shu2019better,jiang2020ltrwes} & 3 \\
AUC and MCC & \cite{laiq2026few} & 1 \\
\bottomrule
\multicolumn{3}{l}{\footnotesize{\textbf{*} ChatGPT model was not reported in the study.}}
\\
\end{tabular}
\end{table}

Bug reports can be classified according to different criteria, such as validity, priority, or whether they represent security-related or non-security-related issues. Consequently, several studies have explored different aspects of bug report classification \cite{laiq2025automatic}. For example, previous work has examined classification based on report validity \cite{laiq2022early,laiq2025comparative} and on differentiating among tasks such as feature requests, questions, and documentation updates \cite{laiq2023intelligent,kallis2019ticket}. Similarly, other studies have focused on grouping or clustering bug reports to identify shared underlying causes \cite{laiq2023data,rahman2020some}.

In this study, we focus specifically on classifying bug reports to distinguish between security-related and non-security-related reports.

Table~\ref{tab:related-work-on-BRC} presents an overview of the identified related work on classifying bug reports as security-related or non-security-related. The studies are grouped by the techniques employed, the types of features used to train a model, and the evaluation strategies and metrics used.

In terms of techniques, as shown in Table~\ref{tab:related-work-on-BRC}, classical ML techniques are predominant. Random Forest, Logistic Regression, Support Vector Machines, and K-Nearest Neighbor are the most commonly applied techniques in these studies. More recent work has expanded the methodological landscape by incorporating deep learning and LLMs, including models based on BERT \cite{cao2022sbrpbert,shu2019better,peters2017text,jiang2020ltrwes}, models based on GPT \cite{yokoyama2024identifying,francca2025gpts}, and few-shot learning approaches such as SetFit \cite{laiq2026few}.

In these studies, various features, including the title, summary, and description of a bug report, have been utilized to train models to classify bug reports to identify security-related bug reports. Most studies only use the summary and description of the bug report. Only a few studies have used additional features such as report timestamps, priority levels, or commit messages.

The evaluation approaches in these studies are highly heterogeneous. As summarized in Table~\ref{tab:related-work-on-BRC}, most studies rely on fixed test sets, although several adopt cross-validation or repeated trials. F1-score, Precision, and Recall are the primary evaluation metrics, while other complementary metrics (e.g., AUC, MCC, or G-measure) are less frequently used. 

The diversity of features, evaluation metrics, and evaluation strategies employed in these studies makes it challenging to compare their findings. In addition, the literature remains fragmented: most studies introduce or refine a particular technique and evaluate it against a limited set of baselines. For example, recent work by Laiq \cite{laiq2026few} introduced a SetFit-based few-shot learning approach to identify security bug reports. However, it only compares SetFit with classical ML techniques. Similarly, the study by Francca et al. \cite{francca2025gpts} compares classical ML techniques with GPT-based models. These two studies do not compare their work with BERT-based models. Similarly, studies that use BERT-based models \cite{soltaniani2025security,cao2022sbrpbert,yokoyama2024identifying}  do not compare their work with SetFit and GPT-based models. Additionally, the experimental settings (e.g., evaluation strategy and metrics) used in these studies are not directly comparable.

These inconsistencies underscore the need for a comprehensive and unified evaluation of existing techniques to facilitate more reliable comparisons and informed adoption decisions.

Motivated by the aforementioned gap, this study systematically and comprehensively evaluates a wide range of techniques for automatically identifying security-related bug reports. We evaluated traditional ML techniques, BERT-based models, OpenAI's GPT-5.2, and SetFit (a state-of-the-art few-shot learning framework) on benchmark datasets (Ambari, Camel, Wicket, and Derby).

\section{Research method}\label{sec:Method}
This study aims to comprehensively evaluate the effectiveness of several promising automatic techniques in identifying security bug reports. To achieve this goal, we posed the following research question: \textbf{\textit{How do automatic techniques perform to identify security bug reports?}} To answer this research question, we perform comparative experiments \cite{alpaydin2020introduction,wohlin2012experimentation}. 

In the following subsections, we describe the key decisions made in designing the experiments, including the selection of automatic techniques, the choice of datasets, the evaluation approach, and the selection of evaluation metrics.

\subsection{Selection of techniques}\label{sec:selection-of-techniques}
As shown in Table~\ref{tab:related-work-on-BRC}, the majority of existing studies on identifying security bug reports are based on classical (traditional) ML techniques.

Similar to previous work (e.g., \cite{francca2025gpts,laiq2026few,shu2019better}), we selected the following three classical ML techniques as baselines: Support Vector Machines, Logistic Regression, and Random Forest. These techniques have been widely used for identifying security bug reports (See Table~\ref{tab:related-work-on-BRC}).

In addition to classical ML techniques, recent work (see Table~\ref{tab:related-work-on-BRC}) has explored the use of LLMs to identify security bug reports. We therefore included the BERT-base and OpenAI's GPT-5.2 models. Beyond existing approaches, this study uses an additional technique (i.e., RoBERTa). Although RoBERTa \cite{liu2019roberta} has achieved state-of-the-art performance in classifying bug reports \cite{laiq2025automatic} into bug and non-bug issues, it has not yet been applied to identify security-related bug reports. Therefore, we include it to assess its performance in this task. Finally, motivated by the limited availability of labeled security bug reports, we incorporate SetFit \cite{koch2015siamese}, a state-of-the-art few-shot learning framework that has previously been studied in isolation (without comparison with other techniques for this task). Together, these techniques enable a unified evaluation across classical ML techniques, LLMs, and the SetFit-based few-shot learning approach, addressing the fragmentation observed in prior work.

\subsection{Selection of datasets}\label{sec:selection-of-dbs}
To evaluate the selected techniques, we utilized four benchmark datasets widely adopted in prior research on security bug report identification \cite{alqahtani2024security,yokoyama2024identifying,francca2025gpts}: Ambari, Camel, Derby, and Wicket. These datasets have been manually labeled Wu et al. \cite{wu2021data}. The validity threats associated with the chosen datasets are discussed in detail in Section \ref{sec:validitythreats}.

An overview of the datasets is provided in Table \ref{tab:datasets}. All four datasets are derived from open-source software projects and were originally compiled and made available by Wu et al. \cite{wu2021data}. Ambari is a Hadoop management platform, Camel is an integration framework, Derby is a relational database management system, and Wicket is a Java-based web application framework. Collectively, these datasets span diverse application domains, helping mitigate project-specific bias and supporting a more generalizable evaluation of the selected techniques.

\begin{table}[ht]
    \centering
\caption{Evaluated datasets of security bug reports (BRs), manually labeled by Wu et al. \cite{wu2021data}}
    \label{tab:datasets}
    \begin{tabular}{cccc}
    \toprule 
\textbf{Project} &  \textbf{Security BRs}  &  \textbf{Non-Security BRs} & \textbf{Total BRs} 
      \\\midrule

Camel &  74 (7.4\%) & 926 & 1000 \\

Ambari & 56 (5.6\%) & 944 & 1000 \\

Derby & 179 (17.9\%) & 821 & 1000 \\

Wicket & 47 (4.7\%) & 953 & 1000 \\
\bottomrule

\end{tabular}
\end{table}

\subsection{Evaluation approach}\label{sec:evalaution-approach}
We use a five-fold cross-validation strategy to evaluate the selected techniques. Compared to evaluation in a fixed train–test split, cross-validation reduces potential sampling bias and provides a more reliable estimate of model performance \cite{witten2002data}. To perform the five-fold cross-validation, we split our datasets into five approximately equal folds. In each iteration, we trained the models using four of these folds and evaluated them on the remaining fold. This procedure was repeated five times to ensure that different subsets were used for testing in each cycle. Ultimately, we averaged the results of all five folds for each technique evaluated.

\textbf{\textit{Evaluation approach for the GPT-based model:}} For GPT-5.2, we use the following approach.
\begin{itemize}
    \item [] \textbf{(a) Zero-shot setting:} In this setting, we use only the test sets for prediction. Table \ref{tab:zero-shot-template} presents a zero-shot prompt template in which the model is instructed to assume the role of a software security expert and to assign a given bug report to one of two predefined categories: security bug or non-security bug. The template provides only task instructions, category definitions, and the target bug report, without including any labeled examples. In addition, the prompt explicitly restricts the model’s output by prohibiting explanations or extraneous text, thus enforcing a concise single-label response. This design assesses the model’s ability to perform the classification task solely based on prior knowledge encoded during pretraining, without requiring task-specific demonstrations.
    \item [] \textbf{(b) Few-shot setting with 2 examples:} In this setting, we include two labeled examples from the training folds as in-context examples, and then evaluate performance on the corresponding test fold. Table \ref{tab:few-shot-template} presents a quick snippet template that enhances the same classification task by using labeled examples derived from the dataset. Before presenting the target bug report, the prompt includes one non-security bug report labeled 0 and one security bug report labeled 1. These examples serve as in-context demonstrations that illustrate both the expected input format and the desired output labels. By providing explicit mappings between bug report descriptions and their corresponding classes, the few-shot template guides the model toward the intended decision boundary and output structure. The final segment of the prompt then requests the classification of the target bug report, leveraging the contextual information provided by the preceding examples. 
    \item [] \textbf{(c) Few-shot setting with 1\% training data}: In this setting, we include 1\% (8 examples) of data from the training folds as in-context examples, and then evaluate performance on the corresponding test fold. The same template shown in Table \ref{tab:few-shot-template} was used for the few-shot setting with 1\% training data; instead of 2 examples, we provided 8 examples for the model.
\end{itemize}
The GPT model was used with the above-mentioned setting, primarily following the recent work by Francca et al. \cite{francca2025gpts}. Although their study focuses on a zero-shot setting, we additionally explored the few-shot setup. Due to the cost associated with these experiments, the number of examples was limited to a maximum of eight. This constraint may affect comparisons with other techniques in this study, which leverage the entire training dataset before testing. However, we consider this a minor limitation, as GPT models are inherently capable of interpreting and generating bug reports and can function as general-purpose assistants for tasks such as text summarization and feature extraction.

%Tables \ref{tab:zero-shot-template} and \ref{tab:few-shot-template} illustrate the prompt templates employed for zero-shot and few-shot learning in the task of classifying software bug reports as security-related or non-security-related.

\begin{table}[h!]
\centering
\caption{Zero-shot prompt template for identifying security-related bug reports, adapted from \cite{yokoyama2024identifying}}
\label{tab:zero-shot-template}
\begin{tabular}{cp{7cm}}
\toprule
\textbf{\#} & \textbf{Prompt content} \\
\midrule
1 & \textbf{\# Task:} You are a software security expert. Classify the given software bug report into exactly one of the two categories below. \\

2 & \textbf{Categories:} (a) \texttt{security bug} (b) \texttt{non-security bug} \\

3 & Do not provide any explanation or additional text. \\
4 & \\
5 & \textbf{\# Description of software bug report:} \\

6 & (Target bug report) \\

7 & \textbf{\# Answer:} \\
\bottomrule

\end{tabular}
\end{table}

\begin{table}[h!]
\centering
\caption{Few-shot prompt template for identifying security-related bug reports, adapted from \cite{yokoyama2024identifying}}
\label{tab:few-shot-template}
\begin{tabular}{cp{7cm}}
\toprule
\textbf{\#} & \textbf{Prompt content} \\
\midrule
1 & \textbf{\# Task:} You are a software security expert. Classify the given software bug report into exactly one of the two categories below. \\

2 & \textbf{Categories:} \\

3 & \texttt{* security bug: 1} \\ 

4 & \texttt{* non-security bug: 0} \\

5 &  \\

6 & \textbf{\# Description of software bug report:} \\

7 & (Non-security bug report from the dataset) \\

8 & \textbf{\# Answer:} \\

9 & 0 \\

10 &  \\

11 & \textbf{\# Description of software bug report:} \\

12 & (Security bug report from the dataset) \\

13 & \textbf{\# Answer:} \\

14 & 1 \\

15 &  \\

16 & \textbf{\# Description of software bug report:} \\

17 & (Target bug report) \\

18 & \textbf{\# Answer:} \\
\bottomrule

\end{tabular}
\end{table}

\subsection{Performance evaluation metrics}\label{sec:evaluation-metrics}
Similar to the existing work (see Table \ref{tab:related-work-on-BRC}), this study measures Precision, Recall, and F1-score to evaluate the performance of the selected techniques.
Recall is calculated as the proportion of correctly identified positive instances among all actual positive instances. Precision is calculated by measuring the proportion of correctly predicted positive instances among all predicted positives. F-score is calculated by computing the harmonic mean of Precision and Recall.

\begin{equation}
Recall = \frac{TP}{TP+FN}
\end{equation}

\begin{equation}
Precision = \frac{TP}{TP+FP}
\end{equation}

\begin{equation}
F-score = 2 * \frac{Precision * Recall}{Precision + Recall}
\end{equation}

\subsection{Data preprocessing} \label{sec:data-preprocessing}

Similar to previous work (see Table \ref{tab:related-work-on-BRC}), we utilize bug report descriptions to train our models. We implement a standard data preprocessing approach for these descriptions, which includes removing special characters, numbers, and hyperlinks.

For our classical ML techniques (Logistic Regression, Random Forest, and Support Vector Machines), we utilized Term Frequency-Inverse Document Frequency (TF-IDF) to convert textual data into sparse matrices.

For the SetFit model, we employed the all-mpnet-base-v2, which is one of the top-performing pre-trained models for generating embeddings \cite{colavito2023few}.

For the RoBERTa and BERT-base models, we used the RoBERTa tokenizer and the BERT-base tokenizer from Hugging Face\footnote{https://huggingface.co/}, respectively.

\subsection{Implementation details}\label{sec:implementation-details}

We used the scikit-learn\footnote{https://scikit-learn.org/} library to implement the selected classical ML techniques. To identify the optimal parameters for these techniques, we performed a grid search using scikit-learn. Initially, we fine-tuned the parameters of each chosen technique through this grid search. After determining the best parameters, we built a model using them.

For implementing SetFit \footnote{https://huggingface.co/docs/setfit/en/index}, RoBERTa, and BERT-base, we follow the official documentation from Hugging Face. To use OpenAI's GPT-5.2 model, we used the OpenAI Python API\footnote{https://pypi.org/project/openai/}.

To perform our experiments, we used a machine equipped with multiple Nvidia H200 NVL GPUs and a couple of AMD EPYC 9565 CPUs, each with cores exceeding 70.

The code and data associated with this work are available at the following links. (1) Dataset provided by Soltaniani et al. \cite{soltaniani2025security}: \url{https://zenodo.org/records/15240583}. (2) Code for this study: \url{https://figshare.com/s/dfedf56689f638c7b35e}.

\section{Results and analysis}\label{sec:Results}
To answer the posed research question (i.e., \textbf{\textit{How do automatic techniques perform to identify security bug reports?}}), we evaluated the performance of the selected techniques, Logistic Regression, Support Vector Machines, Random Forest, RoBERTa, BERT-base, OpenAI's GPT-5.2 (in zero-shot and few-shot settings), and SetFit. These techniques are assessed using four datasets of security bug reports (Camel, Ambari, Derby, and Wicket, see Table \ref{tab:datasets}). In the following, we present the evaluation results of these techniques.

\begin{table*}[!ht]
    \centering
    \small
\caption{Performance of Logistic Regression, Random Forest, Support Vector Machines, RoBERTa, BERT-base, GPT-5.2, and SetFit for identifying security bug reports}
%\vspace{-2mm}
    \label{tab:evaluation-results}
    \begin{tabular}{llcccc}
    \toprule 
\textbf{Dataset} &  \textbf{Technique}  &  \textbf{F-Score} &  \textbf{Recall} &  \textbf{Precision} & \textbf{Accuracy}  
      \\\midrule

\multirow{8}{*}{\centering Camel}
& Logistic Regression & 0.5527 & 0.4857 & 0.6740 & 0.9410 \\

& Support Vector Machines & 0.3433 & 0.4848 & 0.2915 & 0.8620 \\

& Random Forest &  0.3217 & 0.2286 &  0.7306 & 0.9350 \\

& BERT-base & 0.3416 & 0.2695 & 0.6140 & 0.9270 \\

& RoBERTa & 0.5881 & 0.5400 & 0.6581 & 0.9440 \\

& GPT-5.2 (Zero-shot) & 0.1793 & 0.1352 & 0.2683 & 0.9130 \\

& GPT-5.2 (Few-shot, 2 examples) & 0.1318 & 0.0933 & 0.2467 & 0.9160 \\

& GPT-5.2 (Few-shot, 1\%) & 0.1828 & 0.1219 & 0.4143 & 0.9240 \\

& SetFit & \textbf{0.6776} & \textbf{0.5562} & \textbf{0.9118} & \textbf{0.9629} \\
\midrule

\multirow{8}{*}{\centering Ambari}
& Logistic Regression & \textbf{0.4651} & \textbf{0.4818} & \textbf{0.4840} & 0.9350 \\

& Support Vector Machines & 0.3289 & 0.3773 & 0.2995  & 0.9180 \\

& Random Forest & 0.0000 & 0.0000 & 0.0000 & \textbf{0.9440} \\

& BERT-base & 0.2047 & 0.1621 & 0.3133 & 0.9360 \\

& RoBERTa & 0.2489 & 0.1939 & 0.3476 & 0.9370 \\

& GPT-5.2 (Zero-shot) & 0.3227 & 0.2864 & 0.3782 & 0.9340 \\

& GPT-5.2 (Few-shot, 2 examples) & 0.3202 & 0.3364 & 0.3195 & 0.9210 \\

& GPT-5.2 (Few-shot, 1\%) & 0.3438 & 0.3197 & 0.3788 & 0.9330 \\

& SetFit & 0.2340 & 0.1789 & 0.3936 & 0.9380 \\
\midrule

\multirow{8}{*}{\centering Derby}
& Logistic Regression & 0.6850 &  0.6195 & 0.7707 & 0.8990 \\

& Support Vector Machines & 0.5538 & 0.5019 & 0.6712 & 0.8580 \\

& Random Forest & 0.5363 & 0.4190 & 0.7531 & 0.8710 \\

& BERT-base & 0.6793 & 0.6652 & 0.7099 & 0.8910 \\

& RoBERTa & 0.7802 & \textbf{0.7989} & 0.7652 & 0.9180 \\

& GPT-5.2 (Zero-shot) & 0.2674 & 0.1733 & 0.6136 & 0.8360 \\

& GPT-5.2 (Few-shot, 2 examples) & 0.3535 & 0.2459 & 0.6417 & 0.8430 \\

& GPT-5.2 (Few-shot, 1\%) & 0.3106 & 0.2122 & 0.5968 & 0.8350 \\

& SetFit & \textbf{0.7816} & 0.7657 & \textbf{0.8081} & \textbf{0.9240} \\
\midrule

\multirow{8}{*}{\centering Wicket}
& Logistic Regression & 0.5996 & 0.5311 & 0.7057 & 0.9660 \\

& Support Vector Machines & 0.3048 & 0.3600 & 0.2787 & 0.9270 \\

& Random Forest & 0.3508 & 0.2311 & 0.8333  & 0.9610 \\

& BERT-base & 0.3479 & 0.2978 & 0.4444 & 0.9500 \\

& RoBERTa & 0.5997 & 0.4956 & 0.7759 & 0.9670 \\

& GPT-5.2 (Zero-shot) & 0.1329 & 0.1089 & 0.1786 & 0.9340 \\

& GPT-5.2 (Few-shot, 2 examples) & 0.0472 & 0.04222 & 0.05357 & 0.9350 \\

& GPT-5.2 (Few-shot, 1\%) & 0.1358 & 0.1089 & 0.1915 & 0.9340 \\

& SetFit & \textbf{0.7959} & \textbf{0.7133} & \textbf{0.9381} & \textbf{0.9850} \\
\bottomrule

\end{tabular}
\end{table*}

\begin{figure*}[!ht]
    \centering
    
    % Row 1
    \begin{subfigure}[b]{0.48\textwidth}
        \includegraphics[width=\textwidth]{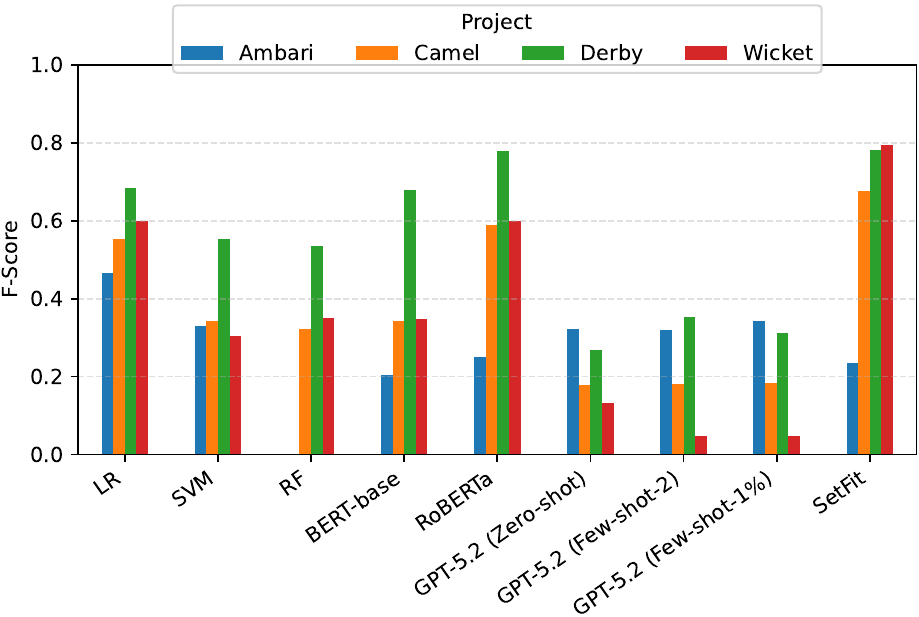}
    \end{subfigure}
    \hfill
    \begin{subfigure}[b]{0.48\textwidth}
        \includegraphics[width=\textwidth]{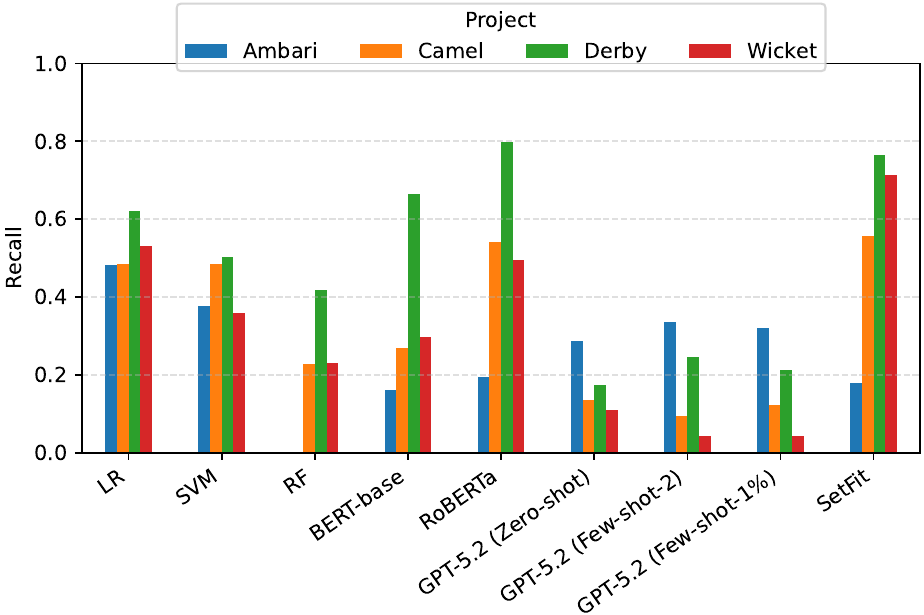}
    \end{subfigure}

    \vspace{1mm}

    % Row 2
    \begin{subfigure}[b]{0.48\textwidth}
        \includegraphics[width=\textwidth]{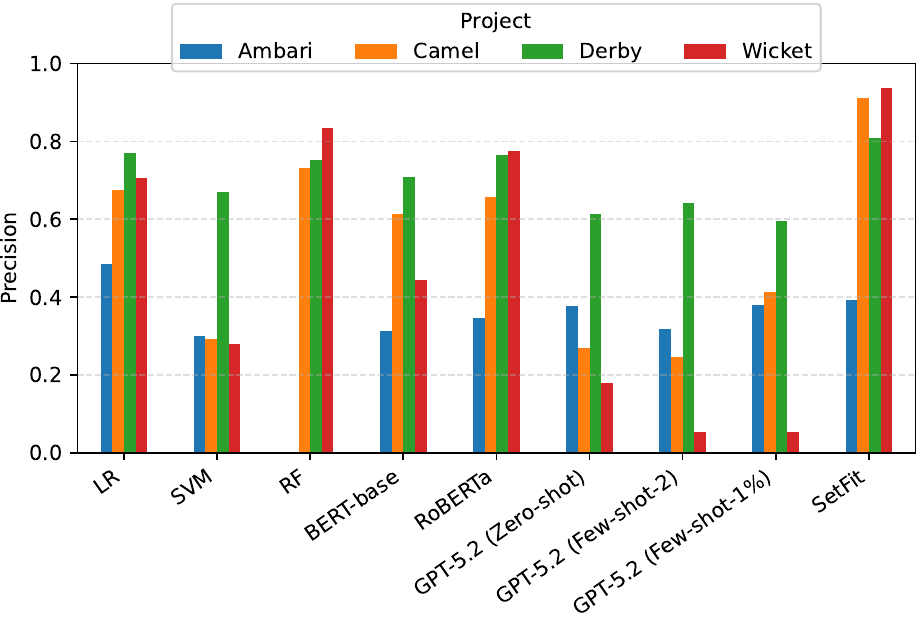}
    \end{subfigure}
    \hfill
    \begin{subfigure}[b]{0.48\textwidth}
        \includegraphics[width=\textwidth]{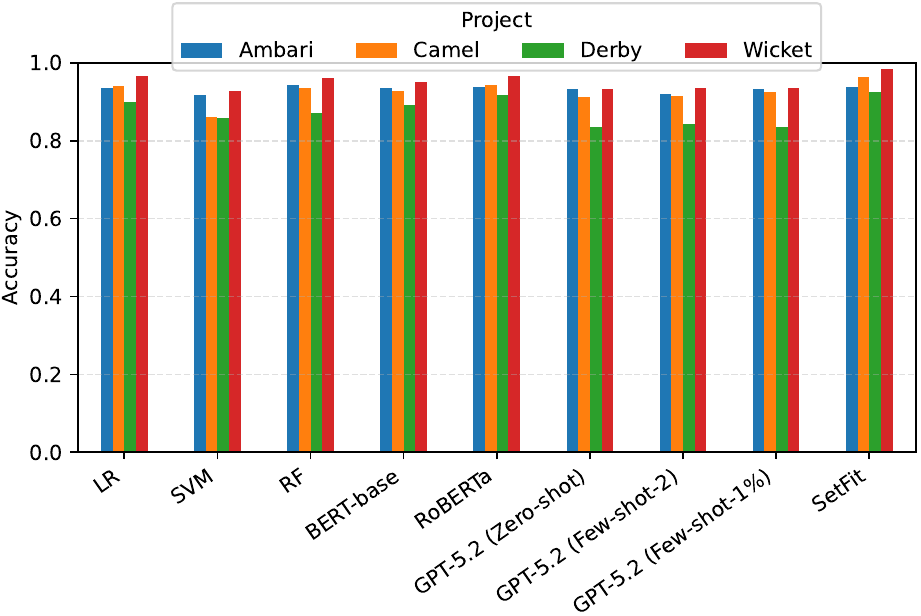}
    \end{subfigure}

    \caption{\centering An overview of the evaluation results. Logistic Regression (LR), Random Forest (RF), Support Vector Machines (SVM), GPT-5.2 (Few-shot-2): GPT-5.2 with 2 examples and GPT-5.2 (Few-shot-1\%): GPT-5.2 with 1\% of the training data}
    \label{fig:evaluation-results}
\end{figure*}

Table \ref{tab:evaluation-results} and Figure \ref{fig:evaluation-results} summarize the performance of all the techniques evaluated. Overall, SetFit achieves the best performance across datasets, except Ambari, particularly in terms of F-score, Recall, Precision, and Accuracy.

\textbf{\textit{Camel project:}}
For the Camel dataset, SetFit clearly outperforms all other techniques, achieving the highest F-score (0.6776), Recall (0.5562), Precision (0.9118), and Accuracy (0.9629). Among the other baselines, RoBERTa is the strongest performer (F-score = 0.5881), followed by Logistic Regression (F-score = 0.5527). Other traditional techniques, Random Forest and Support Vector Machines, perform poorly. GPT-5.2 (both in zero-shot and few-shot settings) shows limited effectiveness, with low F-scores below 0.20, indicating that zero-shot and few-shot prompting are insufficient for accurately identifying security bug reports in this dataset. Overall, SetFit demonstrates a strong balance between Precision and Recall.

\textbf{\textit{Ambari project:}}
The Ambari dataset is particularly challenging for most techniques. Logistic Regression achieves the highest F-score (0.4651) and Recall (0.4818), outperforming all other techniques in this dataset. GPT-5.2 (zero-shot and few-shot) and Support Vector Machines obtain moderate F-scores around 0.32, while BERT-base, RoBERTa, and SetFit perform less effectively in terms of F-score. In particular, Random Forest completely fails to identify any security bug reports. These results confirm the difficulty of identifying security bug reports in the Ambari project, as also observed in prior work \cite{francca2025gpts}.

\textbf{\textit{Derby project:}}
In the Derby dataset, SetFit achieves the best overall performance, with the highest F-score (0.7816), Precision (0.8081), Recall (0.7657), and Accuracy (0.9240). RoBERTa performs competitively, obtaining a similar F-score (0.7802) and slightly higher Recall (0.7989), but with lower Precision. Traditional ML techniques, Logistic Regression, Support Vector Machines, and Random Forest, show reasonable performance but remain clearly below SetFit and RoBERTa. GPT-5.2 again underperforms, particularly in Recall, indicating difficulty in consistently identifying security-related reports without task-specific fine-tuning.

\textbf{\textit{Wicket project:}}
The Wicket dataset further highlights the effectiveness of SetFit, which achieves the highest F-score (0.7959), Recall (0.7133), Precision (0.9381), and Accuracy (0.9850). Logistic Regression and RoBERTa perform similarly in terms of F-score (approximately 0.60), but with noticeably lower Recall. Random Forest exhibits high Precision but extremely low Recall, limiting its usefulness in practice. The variants of GPT-5.2 (zero-shot and few-shot) perform poorly, with F-scores below 0.15, indicating that prompting alone is insufficient for this dataset. SetFit’s strong Recall while maintaining very high Precision is particularly important in security contexts, where missing relevant bug reports can be costly.

\subsection{Generalizability in terms of average F1-score and standard deviation across datasets}
Table~\ref{tab:avg-std-f1} shows the average F1-score and its standard deviation for all techniques in the evaluated datasets. In general, SetFit achieves the best average F1-score (0.6223), indicating the strongest overall performance, though it also exhibits the highest variability (STD = 0.2288). The highest variability in SetFit is primarily due to its poor performance for one of the four datasets (i.e., Ambari). Logistic Regression shows stable performance with low variability (Avg = 0.5756, STD = 0.0795), while RoBERTa achieves competitive results with higher variability. GPT-5.2 configurations yield lower average F1-scores overall, with the few-shot (1\%) setting, based on 8 examples, slightly outperforming the zero-shot and 2-example setups while maintaining relatively low variance.

We note that although these results are based on 5-fold cross-validation, we do not claim statistical significance for the observed differences. In particular, performance differences between top-performing techniques may not be statistically meaningful and should therefore be interpreted with caution.

\begin{table}[!h]
\centering
\caption{Average (Avg) and standard deviation (STD) of F1-score across datasets}
\label{tab:avg-std-f1}
\begin{tabular}{lcc}
\toprule
\textbf{Technique} & \textbf{Avg F1} & \textbf{Std F1} \\
\midrule
Logistic Regression & 0.5756 & 0.0795 \\
Support Vector Machines & 0.3827 & 0.0997 \\
Random Forest & 0.3022 & 0.1930 \\
BERT-base & 0.3934 & 0.1746 \\
RoBERTa & 0.5542 & 0.1920 \\
GPT-5.2 (Zero-shot) & 0.2256 & 0.0740 \\
GPT-5.2 (Few-shot, 2 examples) & 0.2132 & 0.1279 \\
GPT-5.2 (Few-shot, 1\%) & 0.2433 & 0.0864 \\
SetFit & \textbf{0.6223} & \textbf{0.2288} \\
\bottomrule
\end{tabular}
\end{table}

\subsection{Impact of cross-project prediction}\label{sec:cross-project}

\begin{table*}[!ht]
    \centering
    \small
\caption{Cross-project prediction using a leave-one-dataset-out setup: training on three datasets and testing on the remaining dataset}
    \label{tab:evaluation-results-CP}
    \begin{tabular}{llccccc}
    \toprule 
\textbf{Dataset} &  \textbf{Approach} & \textbf{Technique}  &  \textbf{F-Score} &  \textbf{Recall} &  \textbf{Precision} & \textbf{Accuracy}  
      \\\midrule

\multirow{4}{*}{Camel}

& Default & RoBERTa & 0.5881 & 0.5400 & 0.6581 & 0.9440 \\

& Cross-prediction & RoBERTa & \textbf{0.6917} & \textbf{0.6216} & \textbf{0.7797} & \textbf{0.9590} \\ \cmidrule{2-7}

& Default & SetFit & 0.6776 & 0.5562 & \textbf{0.9118} & \textbf{0.9629} \\

& Cross-prediction & SetFit &  \textbf{0.6984} & \textbf{0.5946 }& 0.8462 & 0.9620 \\
\midrule

\multirow{4}{*}{Ambari}

& Default & RoBERTa & 0.2489 & 0.1939 & 0.3476 & \textbf{0.9370} \\

& Cross-prediction & RoBERTa &\textbf{ 0.4071} & \textbf{0.4107} & \textbf{0.4035} & 0.9330 \\ \cmidrule{2-7}

& Default &  SetFit & 0.234 & 0.1789 & 0.3936 & \textbf{0.9380} \\

& Cross-prediction & SetFit & \textbf{0.4248} & \textbf{0.4286} & \textbf{0.4211} & 0.9350 \\
\midrule

\multirow{4}{*}{Derby}

& Default & RoBERTa & \textbf{0.7802} & \textbf{0.7989} & 0.7652 & \textbf{0.9180} \\

& Cross-prediction & RoBERTa & 0.7349 & 0.6816 & \textbf{0.7974 }& 0.9120 \\ \cmidrule{2-7}

& Default &  SetFit & \textbf{0.7816} & 0.7657 & 0.8081 & \textbf{0.9240} \\

& Cross-prediction & SetFit &  0.6232 & 0.4804 & \textbf{0.8866} & 0.8960 \\
\midrule

\multirow{4}{*}{Wicket}

& Default & RoBERTa & \textbf{0.5997 }& 0.4956 & \textbf{0.7759} & \textbf{0.9670} \\

& Cross-prediction & RoBERTa & 0.5902 & \textbf{0.7660} & 0.4800 & 0.9500 \\ \cmidrule{2-7}

& Default &  SetFit & \textbf{0.7959} & \textbf{0.7133} & \textbf{0.9381} & \textbf{0.9850} \\

& Cross-prediction & SetFit & 0.6372 & 0.7660 & 0.5455 & 0.9590 \\
\bottomrule

\end{tabular}
\end{table*}

We also conducted cross-project prediction experiments using SetFit and RoBERTa to assess the impact of transfer learning on the identification of security bug reports. We apply a leave-one-dataset-out strategy: we train a classifier on the full datasets of three projects and test it on the remaining fourth project. This procedure is repeated with each project serving as the target in turn.

Table \ref{tab:evaluation-results-CP} shows the results of the cross-project prediction experiments. Generally, the impact of the prediction between projects varies between datasets and models. 

For \textbf{Camel}, cross-project training improves performance for both approaches. RoBERTa exhibits a notable increase in the F-score from 0.5881 to 0.6917, along with improvements in recall, precision, and accuracy. SetFit also benefits from cross-project prediction, achieving a higher F-score (0.6984 vs. 0.6776) and recall, while maintaining comparably high precision and accuracy.

For \textbf{Ambari}, cross-project prediction yields the most significant gains. Both models show substantial improvements in F-score, increasing from 0.2489 to 0.4071 for RoBERTa and from 0.2340 to 0.4248 for SetFit. These improvements are primarily driven by significant increases in recall, indicating that leveraging data from other projects helps mitigate the limited availability of security bug reports in Ambari.

In contrast, cross-project prediction negatively impacts performance in \textbf{Derby}. Although precision increases slightly for both models, recall drops considerably, resulting in lower F-scores compared to the default within-project setting. This indicates that the characteristics of security bug reports in Derby may be more project-specific and less transferable. 

A similar pattern is observed for \textbf{Wicket}, where cross-project prediction leads to reduced F-scores for both RoBERTa and SetFit. Although recall improves, particularly for RoBERTa, the accompanying decrease in precision leads to overall performance degradation. This suggests that cross-project training introduces more false positives for Wicket.

%Overall, these results indicate that cross-project prediction is most beneficial for projects with limited or highly imbalanced security bug data, such as Ambari and Camel. In contrast, projects like Derby and Wicket benefit more from within-project training due to their stronger project-specific characteristics.

\subsection{Impact of class rebalancing on prediction accuracy}

The datasets used in this study are highly imbalanced (see details in Table~\ref{tab:datasets}). To investigate whether the performance of classical ML techniques can be improved by addressing class imbalance, we applied the widely used Synthetic Minority Over-sampling Technique (SMOTE) \cite{chawla2002smote}. SMOTE was applied after transforming the textual data of bug reports using TF-IDF. Following prior work \cite{shu2021better}, we tuned the (k) parameter of SMOTE and applied the technique only to the training data.

Table~\ref{tab:db-imabalance} presents the results for classical ML techniques under two settings: (i) the \textit{Default} approach without class rebalancing, and (ii) the \textit{Rebalanced} approach using SMOTE. Overall, the results show that SMOTE does not consistently improve performance across datasets or techniques.

For the Camel dataset, SMOTE slightly improves Logistic Regression (F-score increases from 0.4368 to 0.4634), but degrades performance for both SVM and Random Forest. In Ambari, results remain largely poor across all techniques, with SMOTE leading to no improvement for Logistic Regression and SVM, and only a small gain for Random Forest (from 0.0000 to 0.0851).

In the Derby data set, SMOTE provides only a marginal improvement for SVM (from 0.4300 to 0.4339), while reducing performance for Logistic Regression and Random Forest. Similarly, for Wicket, SMOTE improves Random Forest (from 0.2951 to 0.3636), but decreases performance for Logistic Regression and SVM.

Overall, these findings indicate that SMOTE has a limited and inconsistent impact on improving F-score performance for identifying security bug reports. These results may be explained by SMOTE's reliance on nearest-neighbor interpolation, which can be less effective in high-dimensional feature spaces, such as TF-IDF representations of text. We also note that the observed differences between the Default and Rebalanced approaches are relatively small and inconsistent; therefore, we do not claim statistical significance for these differences.

\begin{table}[!h]
    \centering
\caption{Data imbalance handling with SMOTE}
   \small
    \label{tab:db-imabalance}
    \begin{tabular}{llcc}
    \toprule 
\textbf{Dataset} &  \textbf{Approach} & \textbf{Technique}  &  \textbf{F-Score}
\\\midrule

\multirow{6}{*}{Camel}

& Default & Logistic Regression & 0.4368 \\

& Rebalanced & Logistic Regression & \textbf{0.4634 }\\ \cmidrule{2-4}

& Default & SVM & \textbf{0.2500} \\

& Rebalanced & SVM & 0.1935 \\ \cmidrule{2-4}

& Default & Random Forest & \textbf{0.2718} \\

& Rebalanced & Random Forest & 0.2626 \\
\midrule
\multirow{6}{*}{Ambari}

& Default & Logistic Regression & \textbf{0.1429} \\

& Rebalanced & Logistic Regression & 0.0000 \\ \cmidrule{2-4}

& Default & SVM & 0.0000 \\

& Rebalanced & SVM & 0.0000 \\ \cmidrule{2-4}

& Default & Random Forest & 0.0000 \\

& Rebalanced & Random Forest & \textbf{0.0851} \\
\midrule

\multirow{6}{*}{Derby}

& Default & Logistic Regression & \textbf{0.6280} \\

& Rebalanced & Logistic Regression & 0.5980\\ \cmidrule{2-4}

& Default & SVM & 0.4300 \\

& Rebalanced & SVM & \textbf{0.4339} \\ \cmidrule{2-4}

& Default & Random Forest & \textbf{0.4688} \\

& Rebalanced & Random Forest & 0.4000 \\
\midrule

\multirow{6}{*}{Wicket}

& Default & Logistic Regression & \textbf{0.3562} \\

& Rebalanced & Logistic Regression & 0.3385 \\ \cmidrule{2-4}

& Default & SVM & \textbf{0.2308} \\

& Rebalanced & SVM & 0.1667 \\ \cmidrule{2-4}

& Default & Random Forest & 0.2951 \\

& Rebalanced & Random Forest & \textbf{0.3636} \\
\bottomrule

\end{tabular}
\end{table}

\section{Discussion}\label{sec:Discussion}
This study aimed to address the lack of consolidated empirical evidence in prior research on automatically identifying security bug reports by comparing several promising techniques across four benchmark datasets. We evaluated classical ML techniques (Logistic Regression, Support Vector Machines, Random Forest), Bert-based models (RoBERTa and BERT-base), OpenAI's GPT-5.2, and a state-of-the-art few-shot learning framework (SetFit). The results provide insights into how different techniques behave under realistic constraints, such as limited labeled data, severe class imbalance, and cross-project distribution shift. In particular, the findings clarify which approaches are robust when security reports are scarce, when transfer learning is beneficial, and which methodological trade-offs matter most for practical bug triage scenarios.

In evaluations across various projects, few-shot learning with SetFit has proven to be the most effective approach, achieving the highest F1 Scores on three of the four datasets. It demonstrates a favorable balance between precision and recall, which is especially important in security contexts where overlooking vulnerabilities can be costly, while excessive false positives can overwhelm reviewers.  RoBERTa generally performs well and surpasses BERT-base, confirming that stronger pretraining objectives lead to better representations for this task. Meanwhile, classical Logistic Regression remains a surprisingly strong baseline, achieving the best performance on the Ambari dataset. This highlights that simpler models can still be effective, even in situations with extreme class imbalance and limited data. In contrast, the prompt-based OpenAI's GPT-5.2 model performs poorly in both zero-shot and few-shot scenarios, suggesting that prompting is insufficient to capture the nuanced decision boundaries necessary for reliably identifying security issues.

The results of the cross-project prediction reveal that transfer learning is highly dependent on the dataset. Cross-project training improves performance on Camel and Ambari, particularly for Ambari, where the minority class is extremely small, suggesting that exposure to additional security examples can compensate for data scarcity. However, the same strategy degrades performance in Derby and Wicket, primarily due to recall losses or sharp drops in precision. These patterns may indicate significant changes in the distribution between projects, particularly in terms of vocabulary, reporting style, and the composition of security issues. Consequently, “more data” is not inherently beneficial unless it is well aligned with the target project, underscoring the need for careful adaptation and calibration when deploying cross-project models.

From a practical perspective, the results suggest that model selection should be guided by operational priorities. SetFit’s high precision makes it well-suited for triage assistance scenarios where the goal is to flag high-confidence security reports for expedited review, while RoBERTa may be preferable in settings that prioritize recall and can tolerate more false positives. Logistic Regression offers a lightweight, interpretable, and cost-effective option, particularly attractive for organizations with limited computational resources. Importantly, all models are better viewed as decision-support tools rather than fully automated gatekeepers, given the asymmetric costs of errors in security triage.

\section{Threats to validity}\label{sec:validitythreats}
\noindent{\textbf{\textit{Generalizability of the results:}}} To improve the generalizability of our findings, we used multiple datasets (i.e., Camel, Ambari, Derby, and Wicket; see Section \ref{sec:selection-of-dbs} for details). Although these datasets originate from different software projects, they may not fully represent all software development contexts, particularly those that involve proprietary systems.

\noindent{\textbf{\textit{Validity of the evaluation approach:}}} Various evaluation strategies, such as cross-validation and fixed train–test splits, can be adopted to assess the performance of ML techniques. To mitigate potential evaluation bias and improve the reliability of the results, we applied a five-fold cross-validation approach (see Section \ref{sec:evalaution-approach}). This approach helps to reduce experimental bias and improve the reliability of the findings \cite{witten2002data}.

\noindent{\textbf{\textit{Reliability of training data:}}} The quality of training data can introduce validity threats. For example, bug reports may be incorrectly labeled. To address this concern, we relied on four widely used datasets that have been manually validated \cite{wu2021data} for assessing the effectiveness of the selected techniques for identifying security-related bug reports.

Another potential risk related to data is data leakage, i.e., the data used for testing models may have been exposed to the models during their training. This is particularly relevant for pre-trained LLMs such as GPT-5.2 used in this study. However, given the relatively low performance of the GPT-5.2 model, such leakage appears minimal. Therefore, while this risk cannot be entirely dismissed, its practical impact on our findings is likely limited.

\section{Conclusion}\label{sec:Conclusions}
In this study, we conducted a comparative analysis of several automatic techniques to identify security-related bug reports using four benchmark datasets (Ambari, Camel, Derby, and Wicket). We compared traditional ML techniques (Logistic Regression, Support Vector Machines, and Random Forest), BERT-base, RoBERTa, GPT-5.2 (in zero-shot and few-shot settings), and SetFit (a state-of-the-art few-shot learning framework).

Among the evaluated techniques, SetFit demonstrated the overall best performance, achieving the highest F1 scores on three of the four datasets. Within traditional ML techniques, Logistic Regression remained a strong baseline, achieving the best F1-score on the Ambari dataset. In contrast, GPT-5.2 performed poorly in both zero-shot and few-shot settings, suggesting that prompt-only classification is not a reliable substitute for supervised learning in this task.

We further examined cross-project prediction using a leave-one-dataset-out setup for SetFit and RoBERTa. Cross-project training improved performance for Camel and Ambari, particularly for Ambari, where the minority class is very small, indicating that utilizing labeled data from other projects can mitigate data scarcity. However, cross-project training degraded performance in Derby and Wicket, which may indicate differences across projects in reporting style, vocabulary, and the manifestation of security-related issues.

Overall, our findings suggest that SetFit is a strong candidate for practical deployment for identifying security bug reports, particularly when labeled data are scarce. At the same time, the dataset-dependent behavior observed in cross-project prediction highlights the need for project-aware model selection and calibration.

\bibliographystyle{IEEEtran}
\bibliography{paper.bib}

\end{document}